\documentclass[12pt]{article}
\long\def\/*#1*/{}
\usepackage{amsmath}
\usepackage{xcolor}
\usepackage{amssymb}
\usepackage{graphicx}
\usepackage{caption}
\usepackage{float}
\date{}
\begin{document}

\title{Spherical Isentropic Protostars in General Relativity}

\author{Mayer Humi\thanks {e-mail: mhumi@wpi.edu.}\; and John Roumas\thanks{e-mail:roumas@verizon.net}\\
Department of Mathematical Sciences, \\
Worcester Polytechnic Institute, \\
Worcester, MA 01609}

\maketitle

\begin{abstract}

In the process of protostar formation, astrophysical gas clouds undergo 
thermodynamically irreversible processes and emit heat and radiation
to their surroundings. Due the emission of this energy one can envision 
an idealized situation in which the gas entropy remains nearly constant. 
In this setting, we derive in this paper interior solutions to the Einstein 
equations of General Relativity for spheres which consist of isentropic gas. 
To accomplish this objective we derive a single equation for the cumulative 
mass distribution in the protostar. From a solution of this equation 
one can infer readily the coefficients of the metric tensor.
In this paper we present analytic and numerical solutions for the
structure of the isentropic self-gravitating gas. In particular 
we look for solutions in which the mass distribution indicates the presence 
of shells, a possible precursor to solar system formation. Another 
possible physical motivation for this research comes from 
the observation that gamma ray bursts are accompanied by the ejection of 
large amounts of thermodynamically active gas at relativistic velocities. 
Under these conditions it is natural to use the equations of general 
relativity to inquire about the structure of the ejected mass.
\end{abstract}

\thispagestyle{empty}
\newpage
\section{Introduction}
The Einstein equations of General Relativity are highly nonlinear \cite{HR1,HR2}
and their solution presents a challenge that has been addressed by 
many researchers \cite{HR2,HR3}. An early solution of these equations 
is credited to
Schwarzschild for the field exterior to a star \cite{HR4}. However, interior 
solutions (inside space occupied by matter) are especially difficult 
to find due to the fact that the matter energy-momentum tensor is not zero. 
Solutions for this case were derived for static spherical and cylindrical 
symmetry \cite{HR3,HR4,HR5,HR6,HR10b}. In addition various constraints were derived 
on the structure of a spherically symmetric body in static gravitational 
equilibrium \cite{HR7,HR8,HR9,HR10,HR10a}. A conjecture stating that general 
relativistic solutions for shear-free perfect fluids which obey a barotropic 
equation of state are either non-expanding or non-rotating has been 
discussed in a recent review article \cite{HR16}. 
Interior solutions in the presence of anisotropy and other geometries 
were considered also \cite{HR11,HR11a,HR12,HR13}. In addition, interior 
solutions to the Einstein-Maxwell equations have been presented in the 
literature \cite{HR14,HR15}. An exhaustive list of references for exact 
solutions of Einstein equations (up to the year 2009) appears 
in \cite{HR2,HR3}.

In most cases the interior solutions derived in the past considered 
idealized physical conditions such as constant density and pressure 
and ignored thermodynamic irreversible processes 
that might take place in the interior of the (compact) object which lead 
to the emission of radiation and heat. These processes are important 
in the process of protostar formation due to self 
gravitation (prior to nuclear ignition). To take this fact into account 
at least partially, we shall assume that the gas in the interior of these 
objects is isentropic. That is, the entropy produced within the object
(due to the irreversible thermodynamic and turbulent processes taking 
place) is removed by heat and radiation and the gas maintains a constant 
entropy. The same reasoning may apply to mass ejections during gamma ray 
bursts.

For isentropic gas we have the following relationship between
pressure $p$ and density $\rho$
\begin{equation}
\label{1.1}
p=A\rho^{\alpha}
\end{equation}
where $A$ is constant and $\alpha$ is the {\bf isentropy index}.
Two models for $\alpha$ will be considered in this paper, one
with constant $\alpha$ and the other with $\alpha$ as a function of $r$,
the distance from the sphere center.
 
It is our objective in this paper to derive interior solutions for spheres 
which consist of isentropic gas. In particular we shall investigate 
solutions to the Einstein equations which represent spheres in which mass 
is arranged in shells. This structure might then evolve to represent 
the early stages of the process that leads to the formation of a solar 
system. In fact it was Laplace in 1796 who originally put forth the 
hypothesis that planetary systems evolve from a family of isolated rings 
formed from a primitive ``Solar nebula". Such a system of rings 
around a protostar was observed recently by the Atacama Large 
Millimeter/Submillimeter Array in the constellation Taurus.

The plan of the paper is as follows: In Section 2 we review the basic 
theory and equations that govern mass distribution and the components of 
the metric tensor. In Section 3 we present exact, 
approximate and numerical solutions to these equations for spheres made of 
isentropic gas in which the isentropic index is a function of $r$. 
In Section 4 we do the same for spheres with constant isentropic index but 
with $A=A(r)$. We summarize with some conclusions in Section 5.

\setcounter{equation}{0}
\section{Review}

In this section we present a review of the basic theory, following chapter 
$14$ in \cite{HR2}.

The general form of the Einstein equations is
\begin{equation}
\label{2.1}
R_{mn}-\frac{1}{2}g_{mn}R=-\frac{8\pi \kappa}{c^2}T_{mn},\,\,\,m,n=0,1,2,3.
\end{equation}
where $R_{mn}$ and $R$ are respectively the contracted form of the 
Riemann tensor $R_{abcd}$ and the Ricci scalar, 
$$
R_{mn}=R^a_{man},\,\,\, R=R^m_m.
$$
$T_{mn}$ is the matter stress-energy tensor, $\kappa$ is Newton's gravitational 
constant, $c$ is the speed of light in a vacuum and $g_{mn}$ is the 
metric tensor.

The general expression for the stress-energy tensor is 
\begin{equation}
\label{2.2}
T_{mn}=\rho u_mu_n+\frac{p}{c^2}(u_mu_n-g_{mn}),
\end{equation}
where $\rho({\bf x})$ is the proper density of matter and $u_m({\bf x})$ 
is the four vector velocity of the flow.

In the following we shall assume that $\rho=\rho(r)$, $p=p(r)$ and
a metric tensor of the form
\begin{equation}
\label{2.3}
g_{mn}=c^2e^{\nu}dt^2-[e^{\lambda}dr^2+r^2(d\phi^2+\sin^2\phi d\theta^2)].
\end{equation}
where $\lambda=\lambda(r)$, $\nu=\nu(r)$ and $r,\phi,\theta$ are the spherical 
coordinates in 3-space.

When matter is static $u_m=(u_0,0,0,0)$ and $T_{mn}$ takes the following form,
\begin{eqnarray}
\label{2.4}
T_{mn} = \left(\begin{array}{cccc}
\rho e^{\nu} &0&0 &0 \\
0&\frac{p}{c^2}e^{\lambda} &0 &0\\
0 &0 &\frac{p}{c^2}r^2&0 \\
0&0&0&\frac{p}{c^2}r^2\sin^2\phi\\
\end{array}
\right).
\end{eqnarray}
After some algebra \cite{HR2,HR7,HR8} one obtains equations for
$\rho$, $p$, $\lambda$, $\nu$ and $m(r)$ (where $m(r)$ is the total mass
of the sphere up to radius $r$). These are
\begin{equation}
\label{2.5}
\frac{dm}{dr}=Br^2\rho
\end{equation}
\begin{equation}
\label{2.6}
e^{-\lambda}=1-\frac{2m}{r},
\end{equation}
\begin{equation}
\label{2.7}
\frac{e^{\lambda}}{r^2}=\frac{1}{r^2}-
\frac{1}{4}\left[\left(\frac{d\nu}{dr}\right)^2-
\frac{d\nu}{dr}\frac{d\lambda}{dr}\right]
+\frac{1}{2r}\left(\frac{d\nu}{dr}+\frac{d\lambda}{dr}\right)-
\frac{1}{2}\frac{d^2\nu}{dr^2}
\end{equation}
\begin{equation}
\label{2.8}
\frac{C}{c^2}p=\frac{1}{r^2}-e^{-\lambda}\left(\frac{1}{r^2}+
\frac{1}{r}\frac{d\nu}{dr}\right),
\end{equation}
$$
C=-\frac{8\pi\kappa}{c^2}, B=\frac{4\pi\kappa}{c^2},
$$
where $c$ is the speed of light. 
In addition we have the Tolman-Oppenheimer-Volkoff (TOV) equation
which is a consequence of (\ref{2.5})-(\ref{2.8}),
\begin{equation}
\label{2.9}
\frac{1}{c^2}\frac{dp}{dr}=-\frac{m-Cr^3p/2c^2}{r(r-2m)}
\left(\rho+\frac{p}{c^2}\right)
\end{equation}
In the following we normalize $c$ to $1$; $B$ remains $-\frac{C}{2}$.

Assuming that $m(r)$ is known we can solve (\ref{2.7}) algebraically for 
$\lambda$ and substitute the result in (\ref{2.8}) to derive the following 
equation for $\nu$,
\begin{equation}
\label{2.10}
\frac{1}{2}\frac{d^2\nu}{dr^2}+\frac{1}{4}\left(\frac{d\nu}{dr}\right)^2
-\frac{1}{2}\frac{\left(3m-r\frac{dm}{dr}-r\right)\frac{d\nu}{dr}}
{r(2m-r)}-\frac{3m-r\frac{dm}{dr}}{r^2(2m-r)}=0 
\end{equation}
Although this is a nonlinear equation it can be linearized by the
substitution
\begin{equation}
\label{2.11}
\frac{d\nu}{dr}=2\frac{\frac{du}{dr}}{u}=\frac{d\ln(u^2)}{dr}
\end{equation}
which leads to
\begin{equation}
\label{2.12}
\frac{d^2 u}{dr^2}
-\frac{\left(3m-r\frac{dm}{dr}-r\right)}{r(2m-r)}\frac{du}{dr}
-\frac{3m-r\frac{dm}{dr}}{r^2(2m-r)}u=0
\end{equation}
\setcounter{equation}{0}
\section{General Equation for $m(r)$}

Using the equations given in the previous section one can derive 
a single equation for $m(r)$ for a generalized isentropic gas where 
both $A$ and $\alpha$ are functions of $r$
\begin{equation}
\label{7.1}
p=A(r)\rho^{\alpha(r)}.
\end{equation}
To this end we substitute the isentropy relation (\ref{7.1}) in (\ref{2.8})
to obtain
\begin{equation}
\label{7.2}
\rho^{\alpha(r)}=\frac{c^2}{CA(r)}\left\{\frac{1}{r^2}-e^{-\lambda}\left(\frac{1}{r^2}+
\frac{1}{r}\frac{d\nu}{dr}\right)\right\}.
\end{equation}
Using (\ref{2.5}) to substitute for $\rho$ in (\ref{7.2}), normalizing 
$c$ to $1$ and using the fact that $C=-2B$ it follows that
\begin{equation}
\label{7.3}
\left(\frac{\frac{dm(r)}{dr}}{Br^2}\right)^{\alpha(r)}
=-\frac{1}{2BA(r)}\left\{\frac{1}{r^2}-e^{-\lambda}\left(\frac{1}{r^2}+
\frac{1}{r}\frac{d\nu}{dr}\right)\right\}.
\end{equation}
Using (\ref{2.6}) to substitute for $\lambda$ in (\ref{7.3}) and solving the
result for $\frac{d\nu}{dr}$ yields,
\begin{equation}
\label{7.4}
\frac{d\nu}{dr}=-2\frac{\left(\frac{\frac{dm(r)}{dr}}{Br^2}\right)^{\alpha(r)}
BA(r)r^3+m(r)}{r(2m(r)-r)}
\end{equation}
Differentiating this equation to obtain an expression for 
$\frac{d^2\nu}{dr^2}$ and substituting in (\ref{2.10}) leads finally
to the following general equation for $m(r)$  
\begin{eqnarray}
\label{7.5}
&&-2r^{4-2\alpha(r)}B^{1-\alpha(r)}A(r)\alpha(r)(2m(r)-r)^2
\left(\frac{dm(r)}{dr}\right)^{\alpha(r)}\frac{d^2m(r)}{dr^2}+ \\ \notag
&&2m(r)r(2m(r)-r)\left(\frac{dm(r)}{dr}\right)^2-
2m(r)^2(2m(r)-r-1)\frac{dm(r)}{dr}+ \\ \notag
&&2r^{3-2\alpha(r)}B^{1-\alpha(r)}A(r)\{2r^2\alpha(r)+m(r)
[2+8m(r)\alpha(r)+r-2m(r)-8r\alpha(r)]\}
\left(\frac{dm(r)}{dr}\right)^{\alpha(r)+1} \\ \notag
&&+2r^{4-2\alpha(r)}B^{1-\alpha(r)}A(r)(2m(r)-r)
\left(\frac{dm(r)}{dr}\right)^{\alpha(r)+2}+ 
2r^{6-4\alpha(r)}B^{2-2\alpha(r)}A(r)^2
\left(\frac{dm(r)}{dr}\right)^{2\alpha(r)+1} \\ \notag
&&-2r^{4-2\alpha(r)}B^{1-\alpha(r)}(2m(r)-r)^2
\left(\frac{dm(r)}{dr}\right)^{\alpha(r)+1}
\left[\frac{dA(r)}{dr} +
A(r)\ln\left(\frac{\frac{dm(r)}{dr}}{Br^2}\right)\frac{d\alpha(r)}{dr}\right]=0.
\end{eqnarray}
This is a highly nonlinear equation but it simplifies considerably when 
$A(r)$ is a constant or $\alpha(r)$ is an integer. We explore some of 
the numerical solutions of this equation in the next two sections.
A solution of this equation can be used then to compute the metric 
coefficients using (\ref{2.6}) and (\ref{7.4}). With this equation it 
is feasible to investigate the dependence of the mass distribution 
on the parameters $\alpha(r)$ and $A(r)$.

\subsection{Some Analytic Solutions for the Mass Equation}

Although (\ref{7.5}) is highly nolinear, one can obtain analytic
solutions for some predetermined functional values for $m(r)$.

\begin{enumerate}

\item $m(r)=\frac{r}{2},\;\; 0< r < 1$. This ansatz leads to the 
following relation between $\alpha(r)$ and $A(r)$:
\begin{equation}
\label{7.6}
\alpha(r)=\frac{\ln(-2A(r)Br^2)}{\ln(2Br^2)}
\end{equation}
This relation implies that under present assumptions $A(r)$ must be negative.

\item $m(r)=\frac{r^3}{3}$, $A=-1$ yields $\alpha(r)=\frac{\ln(3B)}{\ln B}$ 

\item $m(r)=\frac{r^3}{3}$ and  $\alpha=s$ (where $s$ is a constant)
leads to the following value for $A(r)$:
$$
A(r)=-\frac{B^{s-1}}{3}+\frac{(2r^2-3)B^s}{3Br+C_2B^s\sqrt{2r^2-3}}
$$ 
Here $C_2$ is an integration constant. A similar but algebraically more complicated result can be obtained for $m(r)=Dr^3$ where $D$
is a constant.

\item For $m(r)=\frac{r^3}{3}$ and  $\alpha=r^n$ it follows that
$$
A(r)=\frac{2}{3}\frac{B^{r^n}[(2r^2-3)^{3/2}-C_1(2r-3)(r+1)(2r^2-3)]}
{B(2r+\sqrt{6})(\sqrt{6}-2r)(C_1r+\sqrt{2r^2-3})}
$$
where $C_1$ is an integration constant. A similar result can be obtained for 
$\alpha=Dr^n$ where $D$ is a constant.

\end{enumerate}

It should be observed that the material density $\rho(r)$ for the last three
examples is constant. These examples might therefore represent different routes
for the evolution of a uniform interstellar gas towards the creation of
a protostar (and nuclear ignition). However we were able to obtain also
analytic solutions in terms of hypergeometric and Heun functions for 
$A(r)$ with $\alpha=1$ and $m(r)=r^2$ or $m(r)=r^4$.

\setcounter{equation}{0}

\section{Isentropic Gas Spheres with $p(r)=A\rho(r)^{\alpha(r)}$}

In the following we solve (\ref{2.5}) through (\ref{2.8}) for an isentropic 
gas sphere in which the isentropy index varies with $r$. We discuss three 
examples. The first presents an analytic solution of these equations
while the other two utilize numerical computations.  

\subsection{Isentropic Sphere with Analytic Solution}

When $A(r)$ is a constant and $\alpha$ is a function of $r$ it natural 
to start by choosing a functional form for the density $\rho(r)$ and then 
solve (\ref{2.5}) for $m(r)$. (\ref{2.6}) becomes an algebraic 
equation for $\lambda(r)$ while (\ref{2.7}) is a differential equation 
for $\nu(r)$. Finally, substituting this result in (\ref{7.2}) one 
can compute the isentropy index $\alpha(r)$.

The following illustrates this procedure and leads to an analytic
solution for the metric coefficients.

Consider a sphere of radius $R$ (where $0 < R \le \sqrt{3}$) with the density function
\begin{equation}
\label{3.2}
\rho(r)= \frac{1}{4}\frac{R^2-r^2}{Br^2}
\end{equation}
where $B$ is the constant in (\ref{2.5}). Using (\ref{2.5}) with the initial
condition $m(0)=0$ we then have for $0 \le r \le R$
\begin{equation}
\label{3.3}
m(r)=\frac{R^2r}{4}-\frac{1}{12}r^3.
\end{equation}
Observe that although $\rho(r)$ is singular at $r=0$ the total mass of 
the sphere is finite.

Using (\ref{2.6}) yields
\begin{equation}
\label{3.31}
\lambda(r)=-ln\left(1-\frac{R^2}{2}+\frac{r^2}{6}\right)
\end{equation}

Substituting (\ref{3.3}) in (\ref{2.12}) we obtain a
general solution for $\nu(r)$ which is valid for $R\ne 1$, $R\ne \sqrt{2}$
and $R\ne \sqrt{3}$. It is
\begin{equation}
\label{3.32}
\nu=2\ln(C_1rF(r)^{\omega}+C_2rF(r)^{-\omega})
\end{equation}
where
$$
F(r)=\frac{6-3R^2+\sqrt{6-3R^2}\sqrt{6-3R^2+r^2}}{r},\,\,\,
\omega=\sqrt{\frac{2(R^2-1)}{R^2-2}}.
$$
For R=1 the solution is
\begin{equation}
\label{3.4}
\nu=2\ln\left[r\left(D_1 +D_2 arctanh\sqrt{\frac{3}{3+r^2}}\right)\right]
\end{equation}
At $r=0$ we have $\nu(0)=-\infty$ and the metric is singular at this point.
This reflects the fact that the density function (\ref{3.2}) has 
a singularity at $r=0$ (but the total mass of the sphere is finite). 
To determine the constants $D_1$ and $D_2$ we use the fact that at $R=1$ 
the value of $\nu$ should match the classic Schwarzschild exterior solution
$$
e^{\nu(R)}=1-\frac{2M}{R}
$$
and the pressure (see \ref{2.8}) is zero. These conditions lead to the 
following equations:
\begin{equation}
\label{3.41}
\left(D_1+D_2arctanh\frac{\sqrt{3}}{2}\right)^2-\frac{2}{3}=0,
\end{equation}
\begin{equation}
\label{3.42}
3D_1+3D_2arctanh\frac{\sqrt{3}}{2}-2\sqrt{3}D_2=0.
\end{equation}
The solution of these equations is
$$
D_1=-\frac{\sqrt{2}}{6}\left( 3~arctanh\frac{\sqrt{3}}{2}-2\sqrt{3}\right),\,\,\,
D_2=\frac{\sqrt{2}}{2}.
$$
A plot of $abs(\alpha(r))$ on a semi-log scale is given for this example in Fig. $1$. This graph displays an
unexpected feature which shows that $\alpha$ remains close to zero
except within a region in the middle of the sphere. A possible interpretation 
of this may relate to ongoing thermodynamic processes within the sphere.

For $R=\sqrt{2}$ the differential equation for $\nu$ is
\begin{equation}
\label{3.43}
2\frac{d^2\nu}{dr^2}+\left(\frac{d\nu}{dr}\right)^2+\frac{24}{r^4}=0.
\end{equation}
The solution of this equation is
\begin{equation}
\label{3.44}
\nu=-\ln(24)+2\ln \left[r\left(E_1\sin\frac{\sqrt{6}}{r}+E_2\cos\frac{\sqrt{6}}{r}\right)\right]
\end{equation}
and applying the boundary conditions on $\nu$ and the pressure at $r=\sqrt{2}$
we find that
$$
E_1=2\sin(\sqrt{3}),\,\,\,
E_2=2\cos(\sqrt{3}).
$$
A plot of $\alpha(r)$ exhibits several local spikes in the range 
$0 < r <\sqrt{2}$ but is zero otherwise.

For $\sqrt{2} < R < \sqrt{3}$ the metric coefficient $-e^{\lambda(r)}$
in (\ref{2.3}) becomes
$$
-e^{\lambda(r)}=-\left(1-\frac{R^2}{2}+\frac{r^2}{6}\right)^{-1}.
$$
Therefore for $r< \sqrt{6(\frac{R^2}{2}-1)}$
this metric coefficient is positive and the space has Euclidean structure.
However for $r> \sqrt{6(\frac{R^2}{2}-1)}$ this metric coefficient 
is  negative and the space has a Lorentzian signature. For $R=\sqrt{3}$
the whole interior of the sphere has a Euclidean metric. We consider these 
solutions spurious and have no physical interpretation for their 
peculiar properties at this time.

For $R=\sqrt{3}$  the corresponding differential equation for $\nu$ is
\begin{equation}
\label{3.45}
r^2(6-2r^2)\frac{d^2\nu}{dr^2}+r^2(3-r^2)\left(\frac{d\nu}{dr}\right)^2
-6r\frac{d\nu}{dr}-36=0
\end{equation}
whose general solution is
\begin{equation}
\label{3.46}
\nu=2\ln\left(\frac{A_1(6-r^2)+A_2\sqrt{r^2-3}}{2r}\right).
\end{equation}

\subsection{Infinite Sphere with Density Fluctuations}
 
Consider a sphere of infinite radius with the density function
\begin{equation}
\label{3.11}
\rho=\frac{1}{r^2 k^2}\exp(-\beta r)\sin(kr)^2
\end{equation}
where $\beta$, $k$ are constants and the division by $k^2$ normalizes 
the density to $1$ at $r=0$.

Solving (\ref{2.5}) with the initial condition $m(0)=0$ yields
\begin{equation}
\label{3.12}
m(r)=-\frac{B}{2\beta k^2(\beta^2+4k^2)}
\{e^{-\beta r}[\beta^2+4k^2-\beta^2\cos(2kr)+2\beta k\sin(2kr)]-4k^2\}.
\end{equation}
Observe that although the sphere is assumed to be of infinite radius
the density approaches zero exponentially as $r\rightarrow \infty$
and the total mass of the sphere is finite.

Substituting this result for $m(r)$ into (\ref{2.10}) or (\ref{2.12})
we can solve numerically for $\nu(r)$ and then, 
using (\ref{7.2}), for $\alpha(r)$. Fig. $2$ depicts $\alpha(r)$  
for $\beta=0.001$ and $k=8$. 

\subsection{Finite Sphere with Shell Structure}

We consider a sphere of radius $\pi$ with density function
\begin{equation}
\label{3.21}
\rho=\frac{sin^2(kr)}{k^2r^2}.
\end{equation}
>From (\ref{2.5}) with $m(0)=0$ we then have
\begin{equation}
\label{3.22}
m(r)=\frac{B(2rk-\sin\,2kr)}{4k^3}
\end{equation}
(The total mass $M$ of the sphere is $\frac{B\pi}{2k^2}$).

(\ref{2.10}) was used to solve for $\nu(r)$ numerically 
with the boundary conditions $\nu(0)=0$ and $\nu(\pi)=\ln(1-\frac{2M}{\pi})$ 
so that the value of $\nu(\pi)$ matches that of the Schwarzschild exterior solution
at this point. We then used (\ref{7.2}) to solve for $\alpha$.
Figs. $3,\; 4$ and $5$ depict respectively 
$\rho(r)$, $\nu(r)$ and $\alpha(r)$ for $k=4$.

\subsection{Finite Spheres with Fluctuating $\alpha(r)$}

We considered spheres of radius $1$, total mass of $0.5$, $B=0.1$,
$A=1$ and different fluctuating $\alpha(r)$. Two different
sets of functions were used in these simulations to compute $m(r)$ 
using (\ref{7.5}). In the first set we used the  functions:
\begin{itemize}
\item A.  $\alpha(r)=1+\frac{\sin(4\pi r)}{4}$, 
\item B.  $\alpha(r)=1+\frac{\sin(4\pi r)}{2}$,
\item C.  $\alpha(r)=1+\frac{\sin(8\pi r)}{2}$. 
\end{itemize}
The results of these simulations are presented in Fig. $6$.
We observe that in this figure there are intervals
where $m(r)$ is constant which implies that $\rho(r)\approx 0$ in these
regions. On the other hand a ``step function" in the value of $m(r)$
corresponds to a spike in $\rho(r)$. Therefore
for the functions $B$ and $C$ the mass is distributed in two shells,
one around the ``middle" of the sphere and the other at the boundary. 

For the second set we used the functions
\begin{itemize} 
\item D. $\alpha(r)=1-\frac{r^2}{2}$,
\item E. $\alpha(r)=1-\frac{\sin^2(4\pi r)}{4}$  
\item F. $\alpha(r)=1-\frac{\sin^2(8\pi r)}{4}$. 
\end{itemize}
The results of these simulations are presented in Fig. $7$. 
In this figure the plot for the function $F$ represents a two shell 
structure. However, for the functions $D$ and $E$ there are only ripples in $m(r)$ (which imply the existence of similar ripples in $\rho(r)$).

\setcounter{equation}{0}
\section{Isentropic Spheres with $p(r)=A(r)\rho(r)^{\alpha}$}

In this section we consider isentropic spheres where $\alpha=1$ or $\alpha=2$
with different functions $A(r)$. To solve for the mass 
distribution under these constraints we use the proper reductions of 
(\ref{7.5}).

When $\alpha=1$ (\ref{7.5}) reduces to
\begin{eqnarray}
\label{4.1}
&&-A(r)r^2(2m(r)-r)^2\frac{d^2m(r)}{dr^2}+A(r)r^2(2m(r)-r+A(r))
\left(\frac{dm(r)}{dr}\right)^2 + \\ \notag
&&r\{2A(r)r^2+m(r)[2m(r)(1+3A(r))-r-A(r)(2-7r)]\}\frac{dm(r)}{dr}- \\ \notag
&&m(r)^2(2m(r)-r-1)-r^2(2m(r)-r)^2\frac{dm(r)}{dr}\frac{dA(r)}{dr}=0
\end{eqnarray}
For a sphere of radius  $1$ and $m(1)=0.5$ we display the numerical solution 
of this equation with $A=1$, $A=r$, $A=r^2$ and $A=r^3$ in Fig. $8$. The 
corresponding densities $\rho(r)$ are displayed in Fig. $9$. 
For this set of $A(r)$ functions the total mass is represented
by smooth functions. However there are two peaks in the density, one near $r=0$
and the other at the boundary.

Similarly when $\alpha=2$ we obtain
\begin{eqnarray}
\label{4.2}
&&-2A(r)Br^2(2m(r)-r)^2\frac{dm(r)}{dr}\frac{d^2m(r)}{dr^2}+A(r)^2\left(\frac{dm(r)}{dr}\right)^4 \\ \notag
&&A(r)Br^2(2m(r)-r)\left(\frac{dm(r)}{dr}\right)^3+A(r)Br(4r^2+2m(r)+14m(r)^2-15m(r)r)\left(\frac{dm(r)}{dr}\right)^2+ \\ \notag
&&B^2r^3m(r)(2m(r)-r)\frac{dm(r)}{dr}-B^2r^2m(r)^2(2m(r)-r-1) \\ \notag
&&-Br^2(2m(r)-r)^2\left(\frac{dm(r)}{dr}\right)^2\frac{dA(r)}{dr}=0
\end{eqnarray}

For a sphere of radius  $1$ with $m(1)=0.5$ and $B=0.1$ we display the 
numerical solutions of this equation with $A=-1$, $A=-r$ $A=-r^2$ in Fig. $10$.
In this case $m(r)$ is wavy and as a result frequent fluctuations occur
in the corresponding density function. A shell structure emerges clearly for 
$A(r)=-r^2$.

\section{Conclusions}
In this paper we considered the steady states of a spherical protostar  
or interstellar gas where general relativistic 
considerations have to be taken into account. In addition we considered 
the gas to be isentropic, thereby removing the (implicit or explicit) 
assumption that it is isothermal. Under these assumptions we were able to 
derive a single equation for the total mass of the sphere as a function of 
$r$. From a solution of this equation, the
corresponding metric coefficients may be computed in straightforward fashion.

Our approach was two-pronged. In the first we chose the density distribution
and derived the isentropic index throughout the gas or we let $\alpha$
be a predetermined non-constant function of $r$ and computed $m(r)$.
In the second approach we set the isentropy index to a constant and 
solved the corresponding equation for $m(r)$. In both cases we were able to derive solutions in which the mass is organized in shells.
These solutions represent a new and different class of interior solutions to the
Einstein equations which has not yet been explored in the literature.

\newpage
\begin{figure}[ht]
\centerline{\includegraphics[height=120mm,width=140mm,clip,keepaspectratio]{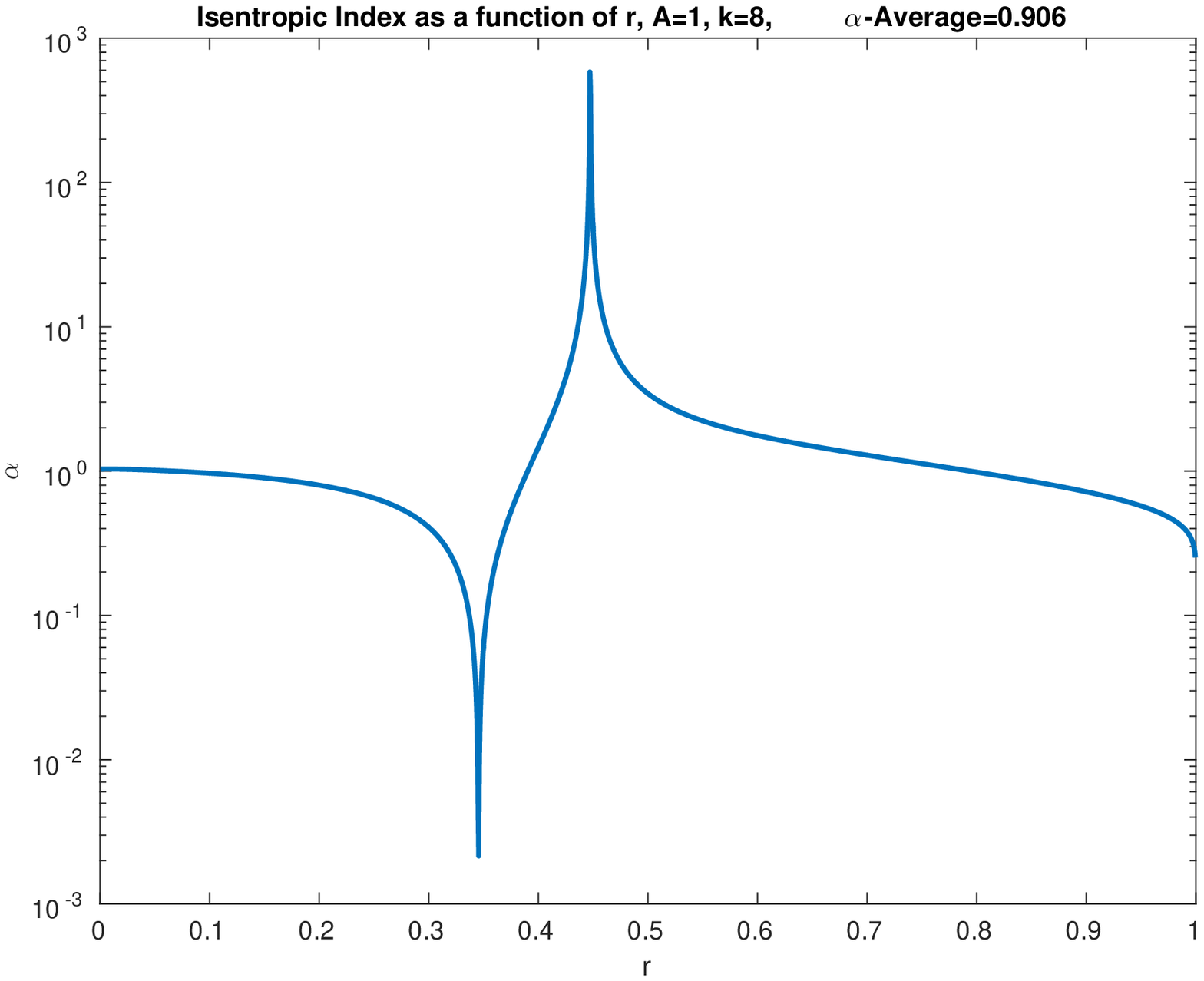}}
\label{Figure 1}
\caption{}
\end{figure}

\newpage
\begin{figure}[ht]
\centerline{\includegraphics[height=120mm,width=140mm,clip,keepaspectratio]{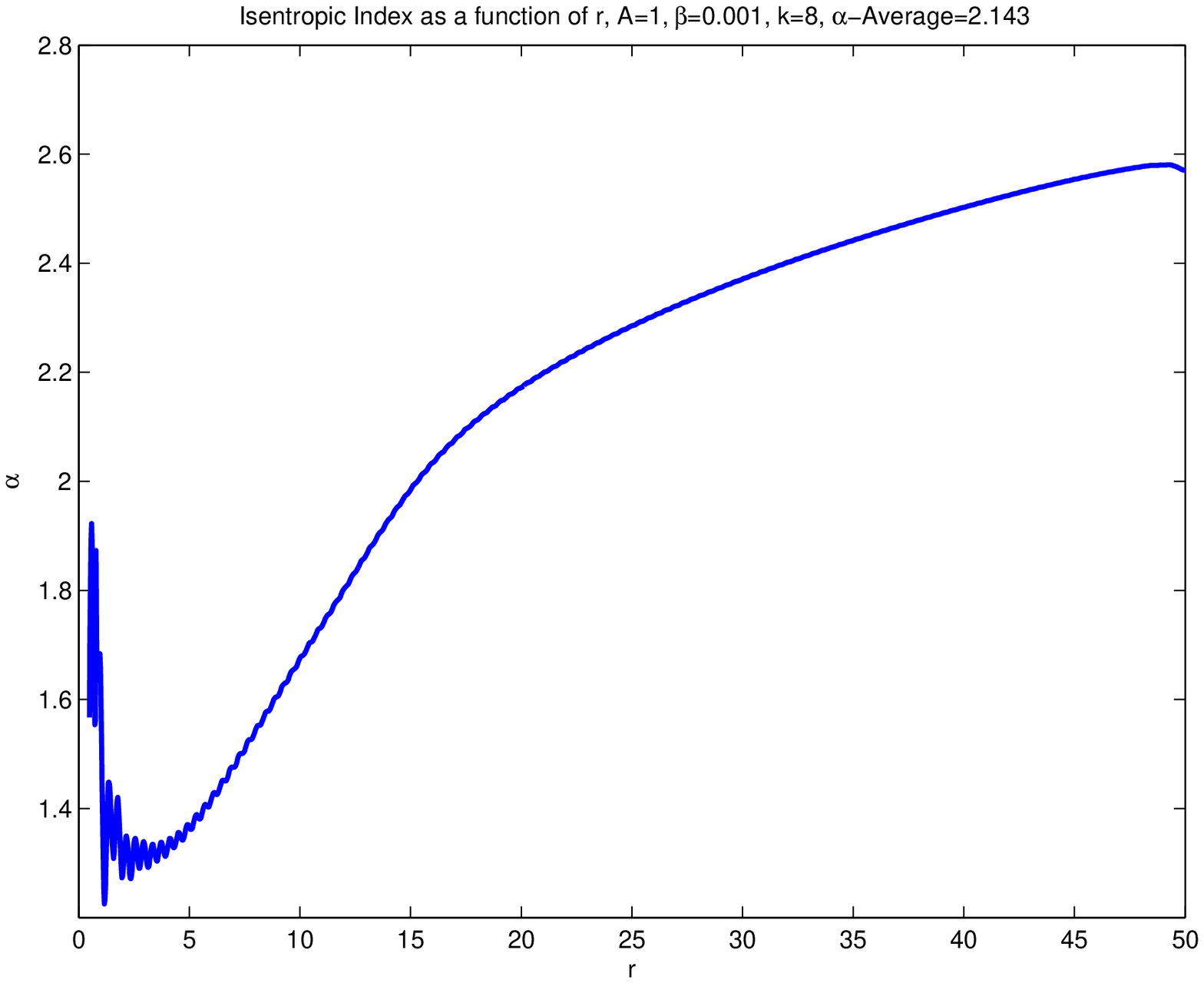}}
\label{Figure 2}
\caption{}
\end{figure}

\newpage
\begin{figure}[ht]
\centerline{\includegraphics[height=100mm,width=120mm,clip,keepaspectratio]{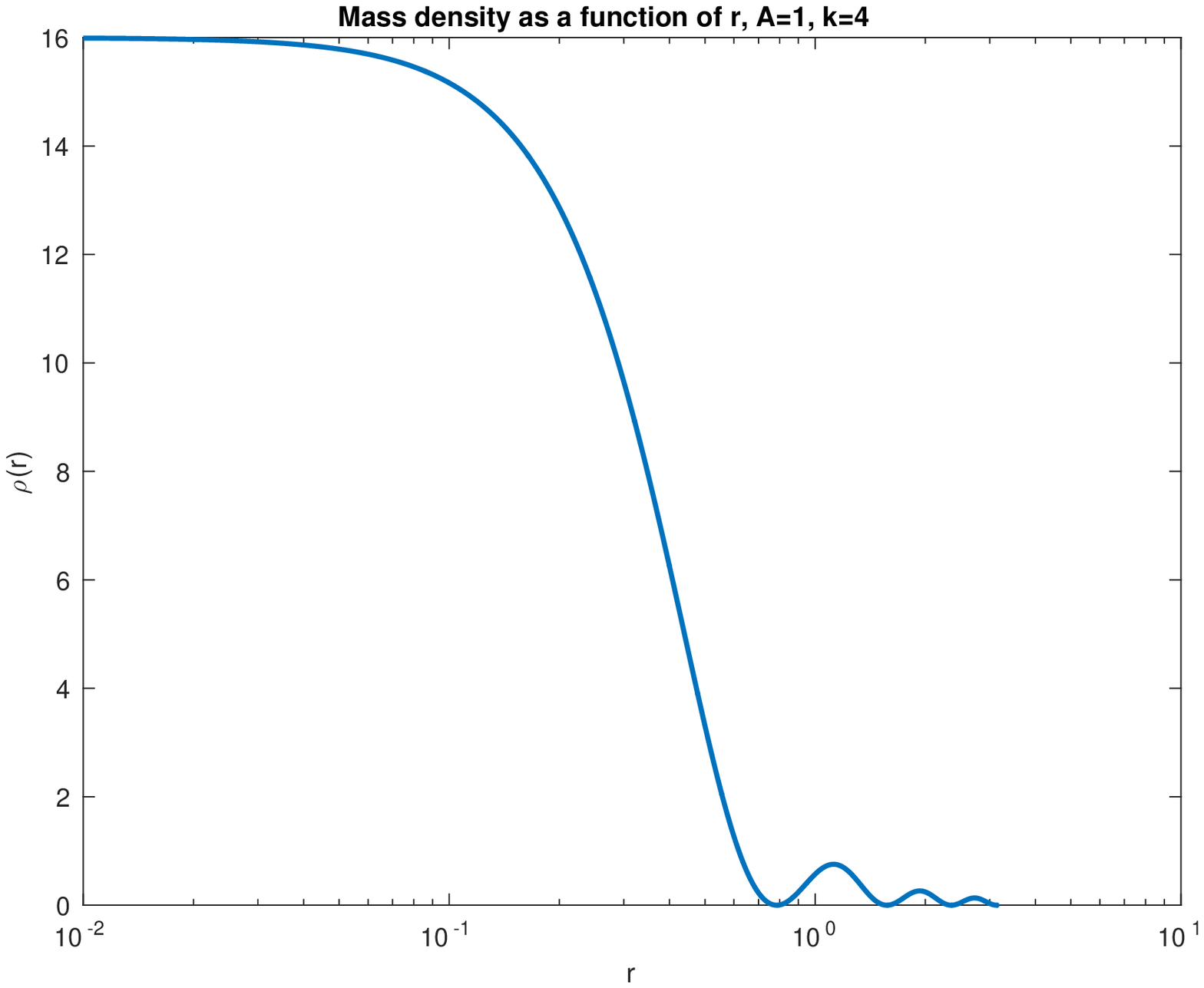}}
\label{Figure 3}
\caption{}
\end{figure}

\newpage
\begin{figure}[ht]
\centerline{\includegraphics[height=100mm,width=120mm,clip,keepaspectratio]{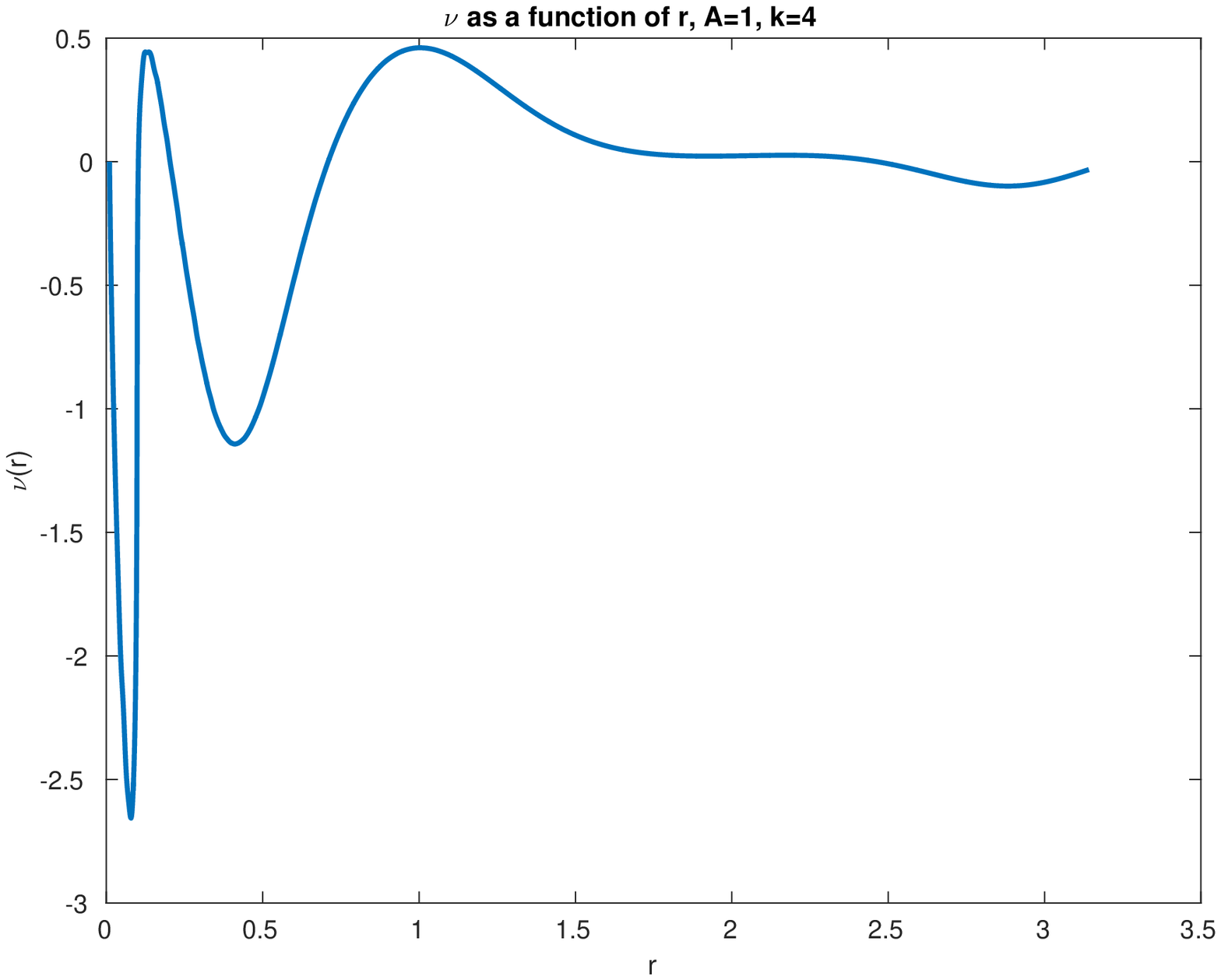}}
\label{Figure 4}
\caption{}
\end{figure}

\newpage
\begin{figure}[ht]
\centerline{\includegraphics[height=120mm,width=140mm,clip,keepaspectratio]{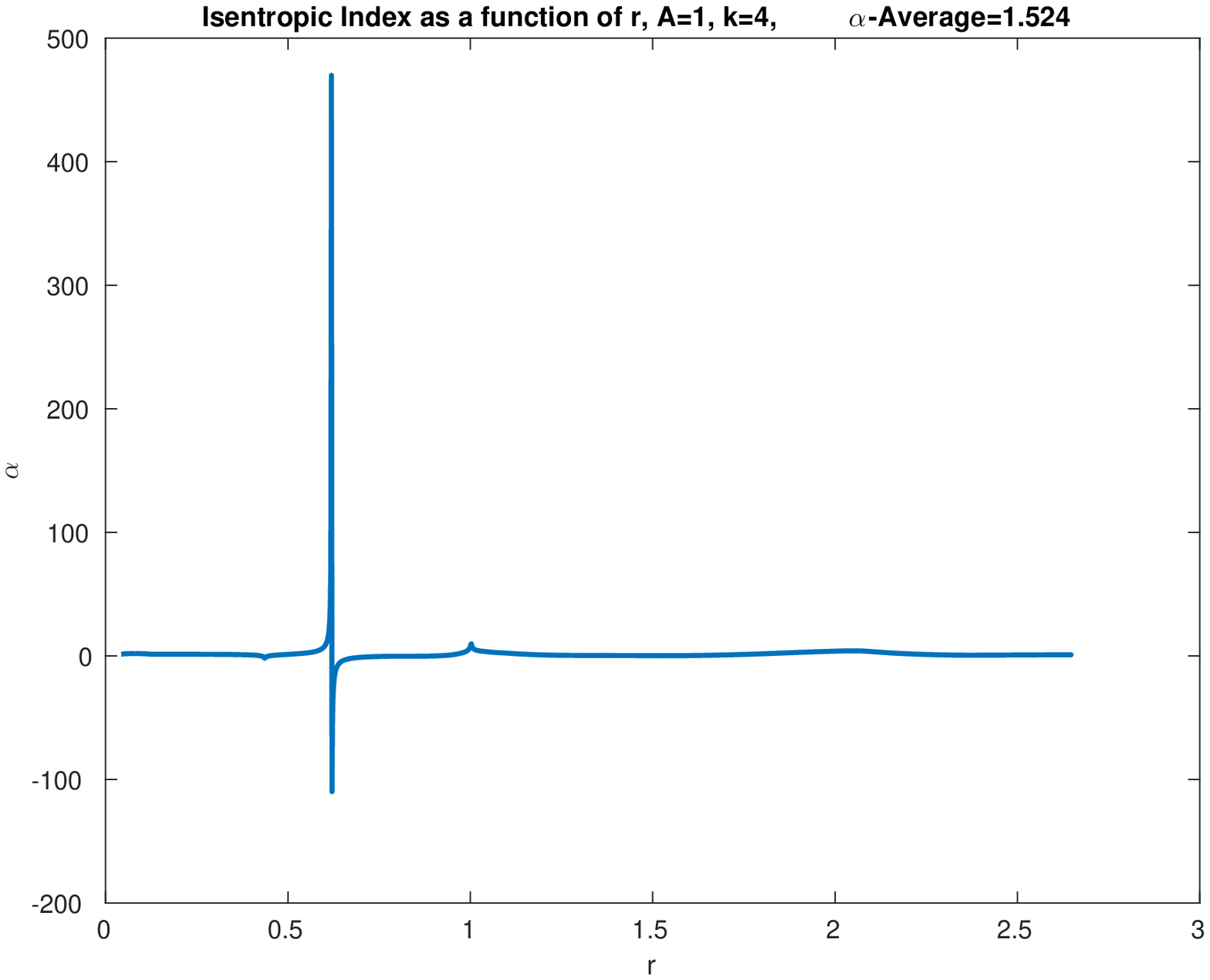}}
\label{Figure 5}
\caption{}
\end{figure}

\newpage
\begin{figure}[ht]
\centerline{\includegraphics[height=120mm,width=140mm,clip,keepaspectratio]{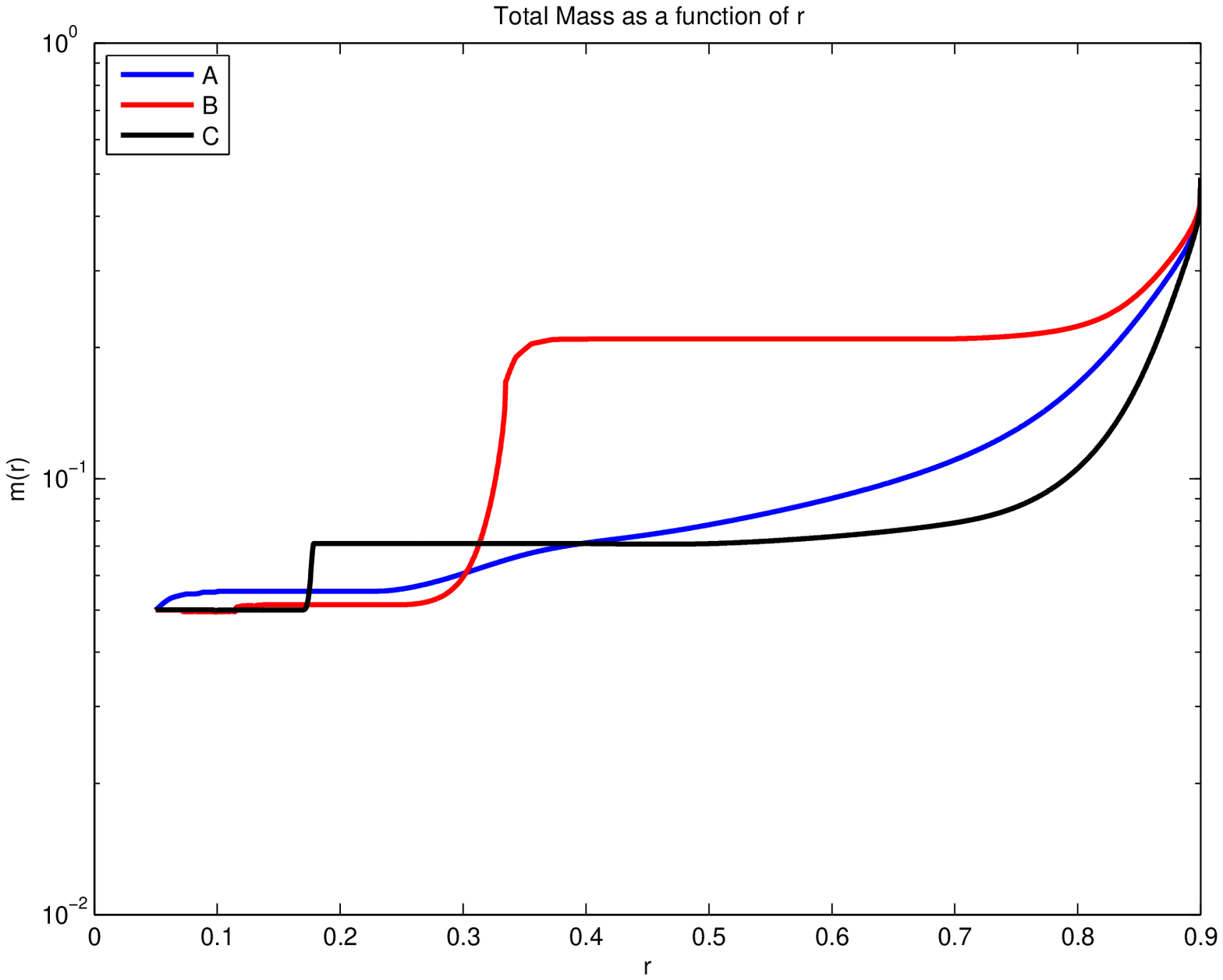}}
\label{Figure 6}
\caption{}
\end{figure}

\newpage
\begin{figure}[ht]
\centerline{\includegraphics[height=120mm,width=140mm,clip,keepaspectratio]{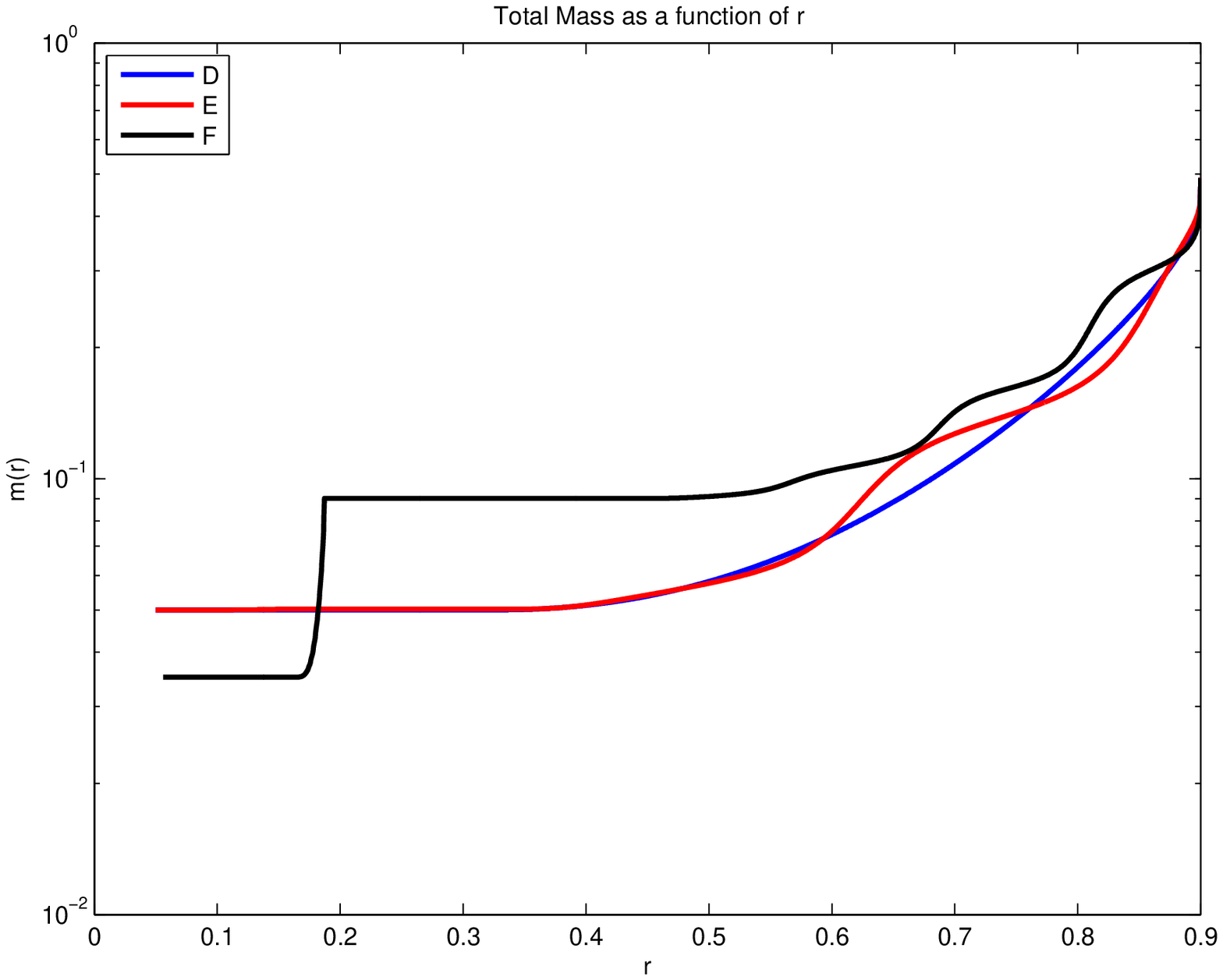}}
\label{Figure 7}
\caption{}
\end{figure}

\newpage
\begin{figure}[ht]
\centerline{\includegraphics[height=120mm,width=140mm,clip,keepaspectratio]{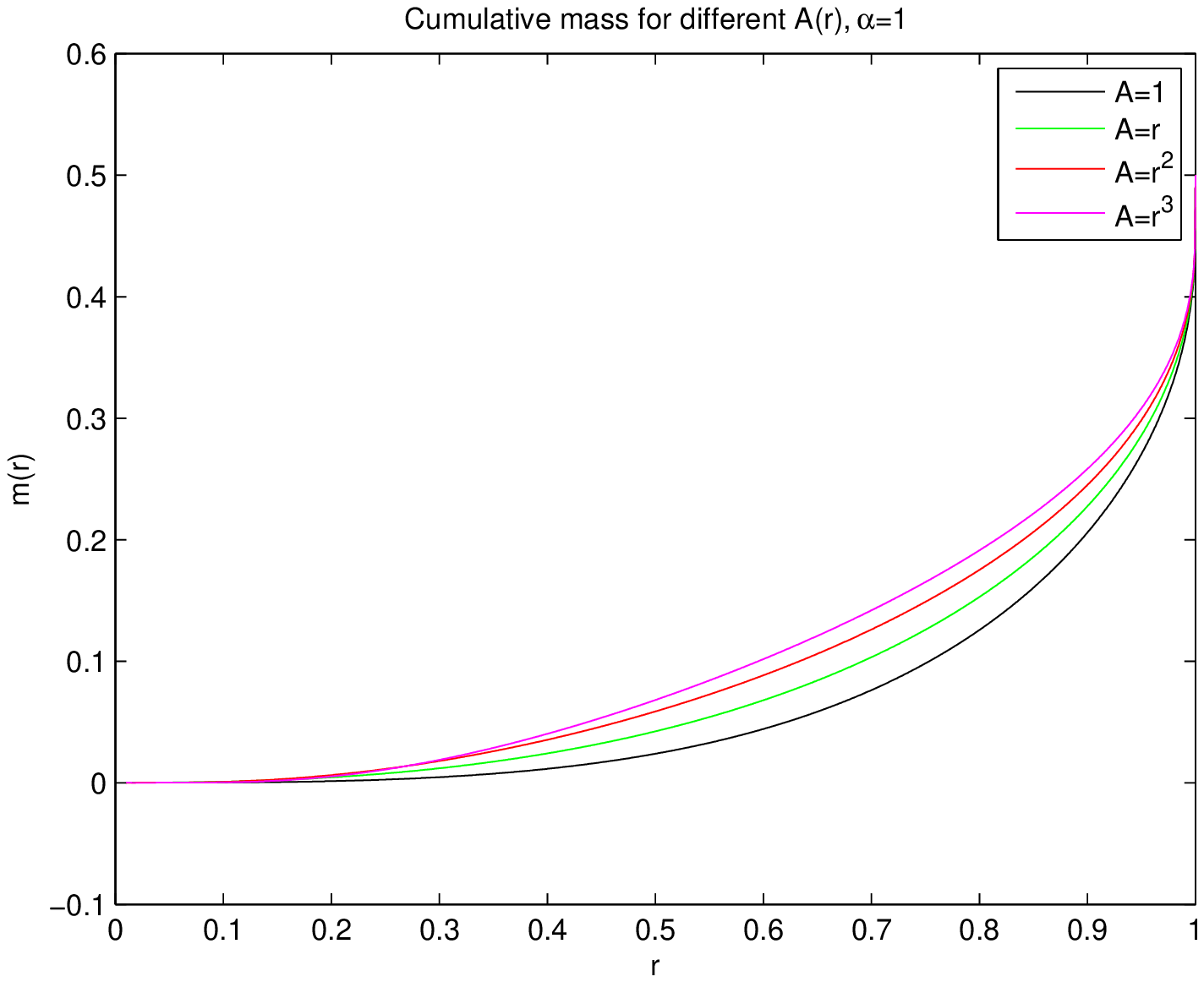}}
\label{Figure 8}
\caption{}
\end{figure}

\newpage
\begin{figure}[ht]
\centerline{\includegraphics[height=120mm,width=140mm,clip,keepaspectratio]{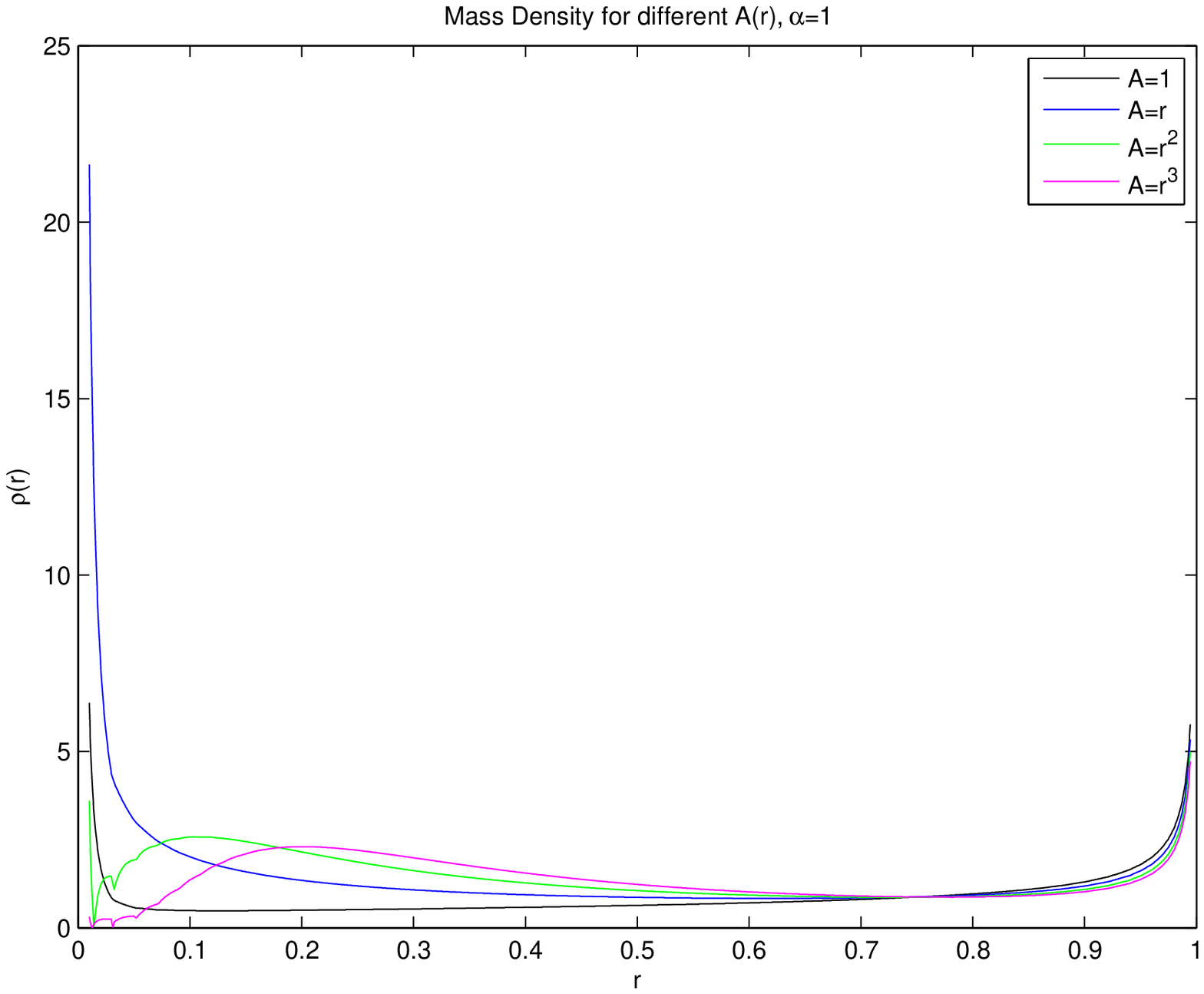}}
\label{Figure 9}
\caption{}
\end{figure}

\newpage
\begin{figure}[ht]
\centerline{\includegraphics[height=120mm,width=140mm,clip,keepaspectratio]{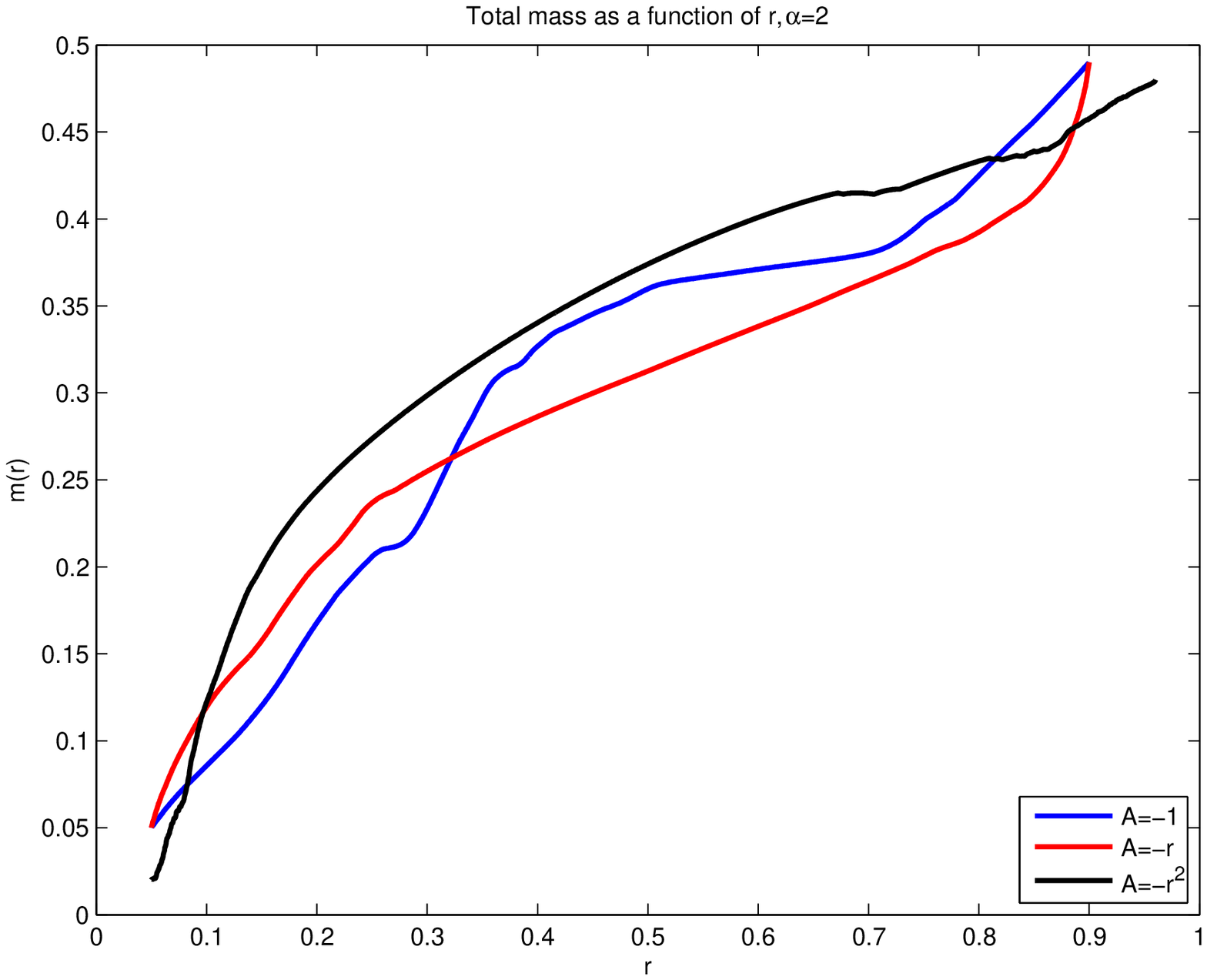}}
\label{Figure 10}
\caption{}
\end{figure}

\end{document}